\documentclass{llncs}
\usepackage{makeidx}  % allows for indexgeneration
\usepackage{amssymb}
\usepackage{amsfonts}
\usepackage{amsmath}
\usepackage{eucal}
\usepackage{graphicx}
\usepackage{float}
\usepackage{booktabs}
\usepackage[ruled, lined, vlined]{algorithm2e}

\begin{document}
\frontmatter          % for the preliminaries
\pagestyle{headings}  % switches on printing of running heads
%\addtocmark{Hamiltonian Mechanics} % additional mark in the TOC
%
\mainmatter              % start of the contributions
\title{Novel Approaches to Accelerating the Convergence Rate of Markov Decision Process for Search Result Diversification}
%
%\titlerunning{Hamiltonian Mechanics}  % abbreviated title (for running head)
\author{Feng Liu\inst{1 \thanks{The work is done when Feng Liu works as an intern in Noah's Ark Lab, Huawei.}} \and Ruiming Tang\inst{2}
\and Xutao Li\inst{1} \and Yunming Ye\inst{1 \thanks{Corresponding author}} \and \\
Huifeng Guo\inst{1} \and Xiuqiang He\inst{2}}
\authorrunning{Feng Liu et al.} % abbreviated author list (for running head)
%Craig Chambers, Kim B. Bruce, and Elisa Bertino}
%
\institute{Shenzhen Key Laboratory of Internet Information Collaboration,\\
Shenzhen Graduate School, Harbin Institute of Technology, Shenzhen, 518055, China,\\
\email{liufeng@stmail.hitsz.edu.cn, lixutao@hitsz.edu.cn, \\
yeyunming@hit.edu.cn, huifengguo@yeah.net}
\and
Noah's Ark Lab, Huawei, China,\\
\email{\{tangruiming, hexiuqiang\}@huawei.com}
}

\maketitle              % typeset the title of the contribution

\begin{abstract}
Recently, some studies have utilized the Markov Decision Process for diversifying (MDP-DIV) the search results in information retrieval. Though promising performances can be delivered, MDP-DIV suffers from a very slow convergence, which hinders its usability in real applications. In this paper, we aim to promote the performance of MDP-DIV by speeding up the convergence rate without much accuracy sacrifice. The slow convergence is incurred by two main reasons: the large action space and data scarcity. On the one hand, the sequential decision making at each position needs to evaluate the query-document relevance for all the candidate set, which results in a huge searching space for MDP; on the other hand, due to the data scarcity, the agent has to proceed more ``trial and error" interactions with the environment. To tackle this problem, we propose MDP-DIV-kNN and MDP-DIV-NTN methods. The MDP-DIV-kNN method adopts a $k$ nearest neighbor strategy, i.e., discarding the $k$ nearest neighbors of the recently-selected action (document), to reduce the diversification searching space. The MDP-DIV-NTN employs a pre-trained diversification neural tensor network (NTN-DIV) as the evaluation model, and combines the results with MDP to produce the final ranking solution. The experiment results demonstrate that the two proposed methods indeed accelerate the convergence rate of the MDP-DIV, which is 3x faster, while the accuracies produced barely degrade, or even are better.

\keywords{search result diversification, Markov decision process, convergence rate}
\end{abstract}
\section{Introduction}

In real web search scenarios, a large number of queries are ambiguous or multi-faceted. For instance, the query ``apple" can be a kind of delicious fruit or the great IT company; the huge vehicle ``rocket'' can also be mentioned as the Houston Rocket basketball team. In order to satisfy the users with different information needs, search result diversification approaches, which provide the search results that covered with a wide range of subtopics for a query, have been widely studied. The approaches work by ranking documents or webpages take both relevance and information novelty (diversification) into considerations.  

A majority of traditional methods for search result diversification are heuristic methods with manually defined functions~\cite{carbonell1998use,dang2012diversity,rafiei2010diversifying,raman2012online,santos2010exploiting,santos2010explicit}. Their key rationale is that the subsequent  document should be ``different" from the ones already ranked. As a representative work, the maximal marginal relevance (MMR)~\cite{carbonell1998use} is proposed to formulate the construction of a diverse ranking as a process of sequential document selection. In MMR, the marginal relevance is defined as a sum of query-document relevance and the maximal document distance as novelty by a predefined document distance function. 

%Another widely used method is xQuAD~\cite{santos2010explicit}, which explicitly models the relationships between the documents retrieved and the possible sub-queries coverage.

Recently, in order to avoid heuristic methods with manually defined evaluation functions, machine learning methods have been proposed and applied to search result diversification~\cite{xia2015learning,xia2016modeling,xu2017directly,yu2017concise,zhu2014learning}. The basic idea is to automatically learn a diverse ranking model from the labeled training data. Typical approaches include the relational learning to rank (R-LTR)~\cite{zhu2014learning} and its variations~\cite{xia2015learning,xia2016modeling,xu2017directly}. In~\cite{xia2015learning,zhu2014learning}, the novelty of a document with respect to the previously selected documents is encoded as a set of handcrafted novelty features. In~\cite{xia2016modeling}, the neural tensor networks are extended to model the novelty among them.

However, all these methods model utility of a candidate document either based on carefully designed heuristics or handcrafted relevance features and novelty features. The utility perceived from the preceding documents is not fully utilized. To avoid this, the latest work for search result diversification, Markov decision process diversification model (MDP-DIV)~\cite{Xia:2017:AMD:3077136.3080775} is proposed, which formalizes the construction of a diverse ranking as a sequential decision making process and models the process with Markov decision process (MDP). Reinforcement learning technique, the policy gradient algorithm of REINFORCE~\cite{sutton1998reinforcement}, is adopted to adjust the model parameters. MDP-DIV outperforms the state-of-the-art baselines on the TREC benchmark datasets. However, its low convergence rate, often requiring tens of thousands iterations to converge, is unacceptable, especially for industrial applications.

In this paper, we aim to promote the performance of MDP-DIV by speeding up the convergence rate and maintaining the accuracy. The primary reasons for low convergence rate are the large action space and data scarcity. On the one hand, the sequential decision making at each position needs to evaluate all the remaining documents of relevance, which forms a huge search space; on the other hand, the data scarcity compels the agent to proceed more ``trial and error" interactions with the environment. To address the problem, we propose MDP-DIV-kNN and MDP-DIV-NTN methods. The MDP-DIV-kNN method adopts a $k$ nearest neighbor strategy to linearly reduce the action space at each position. Specifically, it removes the $k$ nearest neighbors of the recent selected action (document). Different from the MDP-DIV-kNN, the MDP-DIV-NTN employs a pre-trained diversification neural tensor network (NTN-DIV) as the evaluation model, and combines the results with MDP to produce the final ranking list. There are two instantiations of MDP-DIV-NTN. Specifically, the MDP-DIV-NTN(D) method directly filters the pre-ranked list; while the MDP-DIV-NTN(E) method sequentially models the novelty of candidate document with respect to previously selected documents. The main contributions of this paper can be summarized as follows:
\begin{itemize}
  \item We analyze the reasons for the slow convergence of MDP-DIV, and find that it is mainly due to the large action space and the data scarcity.
  \item We propose the MDP-DIV-kNN and MDP-DIV-NTN methods, which can promote the convergence rate while maintaining the accuracy of MDP-DIV for search result diversification.
  \item Extensive experiments are carried out on 09-12 TREC benchmark datasets, and the results demonstrate the proposed methods indeed fasten MDP-DIV and outperforms the state-of-the-art competitors.
\end{itemize}

The remainder of the paper is structured as follows. In Section 2, we briefly review the related works. In Section 3, the Markov decision process, MDP-DIV, and NTN-DIV are introduced as preliminaries. The proposed methods are presented in Section 4. Experimental results are provided in Section 5 to demonstrate the effectiveness of the proposed methods.

\section{Related Work}

\subsection{Search Result Diversification}
One of the key problems in search result diversification is the diverse ranking. Formalizing the construction of diverse ranking as a process of sequential document selection is a common practice. This ranking strategy provides us a more rational way to model the utility of a candidate document which not only depends on the document itself but also the preceding documents. Existing approaches can be classified into two categories, namely heuristic methods~\cite{carbonell1998use,dang2012diversity,guo2010probabilistic,he2012combining,santos2010explicit} and machine learning methods~\cite{xia2015learning,xia2016modeling,Xia:2017:AMD:3077136.3080775,xu2017directly,zhu2014learning}. 

The representative work in the first kind is the maximal marginal relevance (MMR)~\cite{carbonell1998use} criterion to guide the design of diverse ranking models. In MMR, the sequential document selection is based on the marginal relevance score, which is a linear combination of query-document relevance score and document novelty score. A variation of MMR is the probabilistic latent MMR model proposed by Guo and Scanner~\cite{guo2010probabilistic}. PM-2~\cite{dang2012diversity} tackles the problem from the perspective of proportionality. xQuAD~\cite{santos2010explicit} explicitly models the relationships between the documents retrieved for the query and the possible sub-queries coverage. The authors in ~\cite{he2012combining} propose to combine the implicit and explicit topic representations for constructing diverse ranking. All these methods model the utility of candidate document based on carefully designed heuristics with manually defined evaluation functions. However, it is hard to design an unified similarity function for different tasks.

%Hu et al.~\cite{hu2015search} propose a diversification framework that explicitly leverages the hierarchical intents of queries for diverse ranking.

Recently, machine learning approaches have been proposed for search result diversification issue. The ranking score for diverse ranking is based on a linear combination of relevance features and novelty features, and the parameters can be automatically adjusted from the training data. Zhu et al.~\cite{zhu2014learning} propose the relational learning to rank (R-LTR) framework by optimizing the objective function to construct the diverse ranking model. With different definitions of the objective functions and optimization techniques, different diverse ranking algorithms have been proposed~\cite{xia2015learning,xia2016modeling,xu2017directly}. Xia et al.~\cite{xia2015learning} learn a maximal marginal relevance model via directly optimizing diversity evaluation measures. The authors in ~\cite{xia2016modeling} utilize the neural tensor network to model the novelty relations. To avoid the handcrafted features and fully utilize the utility in preceding documents, Xia et al.~\cite{Xia:2017:AMD:3077136.3080775} propose to adapt reinforcement learning techniques to formalize the diverse ranking as a process of sequential decision making which can be modeled with MDP, where the parameters can be trained by policy gradient algorithm of REINFORCE~\cite{sutton1998reinforcement}. 

\subsection{Reinforcement Learning for Information Retrieval}
Reinforcement learning (RL) techniques are widely used in information retrieval (IR) applications. The aforementioned MDP for diverse ranking in~\cite{Xia:2017:AMD:3077136.3080775} is a representative work in this kind. What's more, MDP also can be extended to learning to rank problems~\cite{Wei:2017:RLR:3077136.3080685}, in which the proposed MDPRank model utilizes the MDP to directly optimize the NDCG at all ranking positions. Wang et al.~\cite{Wang:2017:IMG:3077136.3080786} propose a game theoretical minimax game to iteratively optimize the generative retrieval and discriminative retrieval models, in which the generative retrieval model is optimized by the policy gradient algorithm of REINFORCE. In session search, Luo et al.~\cite{luo2014win} propose to utilize the partially observed Markov decision process (POMDP) to model session search as a dual-agent stochastic game for constructing a win-win search framework. The authors in~\cite{zhang2014pomdp} propose to utilize the log-based document re-ranking, which is modeled as a POMDP to improve the ranking performance. Moreover, RL techniques are also utilized in recommender systems. For instance, Guy et al.~\cite{shani2005mdp} designed a MDP based recommender system which employs a strong initial model to converge quickly. The multi-armed bandits technique is also utilized for diverse ranking~\cite{radlinski2008learning}. Lu and Yang~\cite{lu2016partially} propose a neural-optimized POMDP model for building a collaborative filtering recommender system. 

Recent advances in reinforcement learning techniques make the research in IR one step further, and promising performances are delivered, such as MDP-DIV, MDPRank, etc. However, MDP-DIV suffers from a very slow convergence, which hinders the usability in real applications. In this paper, we aim to promote the performance of MDP-DIV by speeding up the convergence rate without much accuracy sacrifice.

\section{Preliminaries}

\subsection{Markov Decision Process}
The search result diversification issue considered in this paper could be formulated with a continuous state Markov decision process (MDP)~\cite{puterman2014markov,sutton1998reinforcement} which is usually utilized for sequential decision making. An MDP is comprised of states, actions, rewards, policy, and transition, and can be represented by a tuple $\langle S, A, T, R, \pi \rangle$

\textbf{States} S is a set of states. In~\cite{Xia:2017:AMD:3077136.3080775}, states can be defined as tuples consisting of preceding ranked documents, candidate documents, and the utility that the agent perceives from the preceding documents as well as the query.

\textbf{Actions} A is a discrete set of actions that an agent can take. The possible actions at each time step depend on the current state $s$, denoted as $A(s)$.

\textbf{Transition} T is the state transition function $s_{t+1}=T(s_t,a_t)$ which maps a state $s_t$ into a new state $s_{t+1}$ in response to the selected action $a_t$.

\textbf{Reward} $r=R(s,a)$ is the immediate reward, also known as reinforcement. It gives the agent an immediate reward when taking action $a$ under state $s$.

\textbf{Policy} $\pi (a|s)$ describes the behaviors of an agent which is a sequence mapping from states to actions. Generally speaking, $\pi$ is optimized to decide how to move around in the state space to achieve the optimal long-term discounted reward $\sum\nolimits_{t = 1}^\infty \gamma^t {{r_t}}$.

The agent interacts with the environment at each time step. For instance, at time step t, the agent receives the environment's state $s_t \in S$, and then selects an action $a_t \in A(s_t)$ based on the current state $s_t$, where $A(s_t)$ is the set of actions available under state $s_t$. As a consequence of the action taken, the agent receives a numerical reward $r_{t+1} \in \mathbb{R}$ and the state changes to $s_{t+1}=T(s_t,a_t)$ simultaneously in the next time step. 

%Figure 1 illustrate the agent-environment interaction in MDP.

%\begin{figure}[ht]
%  \centering
%  \includegraphics[height=3cm, width=0.6\textwidth]{./RL.pdf}
%\caption{The agent-environment interaction in MDP}
%\label{fig:RL}
%\end{figure}

\subsection{MDP-DIV}
MDP-DIV is proposed by Xia et al.~\cite{Xia:2017:AMD:3077136.3080775}, which is the latest and the first approach that utilizes the reinforcement learning techniques for search result diversification. The construction of diverse ranking is formalized as a process of sequential decision making, which is modeled with a continuous state Markov decision process (MDP). The user's perceived utility can be treated as a part of its MDP state. 

%In MDP-DIV, the ranking of documents is formalized as a sequence of $M$ decisions and each action corresponds to selecting one document for a specific position from the candidate set.

%And $s_t$ can be represented by a tuple $s_t=[\CMcal{Z}_t,X_t,h_t]$, where $\CMcal{Z}_t$ is the sequence of $t$ preceding documents, $X_t$ is set of candidate documents and $h_t \in \mathbb{R}^k$ is a vector that encodes the user perceived utility from preceding documents in $\CMcal{Z}^t$ as well as the information need on query $q$.

More specifically, at time step $t$, the agent receives the environment's state $s_t$ which models the user's dynamic state on the perceived utility, starting from the first ranking position. Based on the received state, the agent chooses an action $a_t \in A(s_t)$ depending on the policy that the agent has learned recently. The policy in MDP-DIV is formulated as a {\it softmax} type of function that maps from the current state to a probability distribution of selecting each possible actions. According to the selected action (document), the user perceives some additional utility, also known as the immediate reward, from the recently-selected document. Here the reward is defined as the quality improvement of the selected documents in terms of $\alpha$-DCG or Subtopic recall (S-recall), which are two widely used metrics in search result diversification. Then the system transits to a new state. The transition function, which maps old state and the selected document to a new state, is implemented in a recurrent manner. Reinforcement Learning techniques, the policy gradient algorithm of REINFORCE~\cite{sutton1998reinforcement}, is adopted to coordinate the model parameters for the sake of maximizing the expected long-term discounted rewards. 

The end-to-end MDP-DIV model unifies the relevance and novelty as the criterion for selecting documents which directly optimizes a diversity evaluation measure, and outperforms the state-of-the-art baselines on the TREC benchmark datasets. However, the low convergence rate of needing tens of thousands iterations in the training phase is indeed unacceptable, especially for industrial applications. The reasons are two fold: (i) In the training stage, for decision making at each ranking position, the agent has to go through the whole remaining candidate set which introduces high computational complexity. Suppose we are given $N$ training queries, and each query is associated with a set of $M$ retrieved documents\footnote{For the ease of explaination, we suppose each query is associated with the same number of documents.}. The diverse ranking process will cost $N({\textstyle \frac{1}{2}}M(M + 1))$ times of query-document relevance evaluations for just one iteration. Moreover, the reinforcement learning process often needs large numbers of iterations to converge. Therefore, it is really a catastrophe if we are unfortunately facing to a large discrete action space, i.e. M is large; (ii) The retrieved documents are too scarce to train, which means that the agent has to proceed more ``trial and error'' interactions with the environment. For instance, more than 70\% of data utilized in MDP-DIV are not labeled (i.e., no subtopics is contained). Worse still, some queries are associated with completely irrelevant (unlabeled) documents. 

%Moreover, majority of the labeled data only has one label, and this is much less than the size of label space for a query. 

%Therefore, these probably are the reasons why the MDP-DIV model costs almost $65$ hours to train in a well performed intel$^{\circledR}$ Xeon$^{\circledR}$ Processor E5 V4 server.

\subsection{NTN-DIV}
The NTN-DIV model is proposed by Xia et al.~\cite{xia2016modeling} that models document novelty with neural tensor networks. Intuitively, the neural tensor networks model the relationships between two entities with a bilinear tensor product. This idea could be naturally extended to model the novelty relation of a document with respect to the other documents for search result diversification. Suppose we are given a set of $M$ candidate documents ${\rm{X}} = \{ {d_j}\} _{j = 1}^M$, where each document is characterized with its preliminary representation with embedding models, such as the doc2vec model. The novelty score of a candidate document $d \in X$ with its preliminary representation $v$, and a set of ranked documents  $S \in X\backslash \{ d\}$ with their representations $\{ {v_1},...,{v_{\left| S \right|}}\}$ can be defined as a neural tensor network with $z$ hidden slices. The ranking function can be defined in Eq.(\ref{equ:equ1}):

\begin{equation}\label{equ:equ1}
{f_n}(v,S) = \omega^T v + {\mu ^T}\max \{ \tanh ({v^T}{W^{[1:z]}}[{v_1},...,{v_{\left| S \right|}}])\}
\end{equation}

\noindent where the first term is the relevance score\footnote{In order to learn end-to-end, we use the embedding features instead of handcrafted relevance features.}, and $\omega$ weights the embedding feature $v$. The second term is the novelty score computed by neural tensor network. Specially, $W^{[1:z]}$, a $z$ dimensional three-way tensor, represents the relationship of the documents, where $W_{ijk}$ stands for the $k$-th feature of relationship between documents $d_i$ and $d_j$. And $\mu$ weights the importance of the slices of the tensor. The primary merit of using neural tensor network to model the document novelty is that the tensor can relate the candidate document and the selected documents multiplicatively, instead of only going through a predefined similarity function or through a linear combination of novelty features. To the best of our knowledge, the NTN-DIV model is the latest and the best approach for search result diversification except for MDP-DIV.
%\begin{equation}\label{equ:equ1}
%\pi ({a_t}|[{\CMcal{Z}_t},{X_t},{h_t}]) = \frac{{\exp \{ x_{m({a_t})}^TU{h_t}\} }}{Z}
%Z = \sum\limits_{a \in A({s_t})} {\exp \{ x_{m(a)}^TU{h_t}\} }
%\end{equation}
\section{Methodology}

As aforementioned that large action space and data scarcity will lead to low convergence rate, in this paper, we propose two kinds of strategies to deal with this issue. The first one is the $k$ nearest neighbor strategy, which discards the $k$ nearest neighbors of the recently-selected action (document); The second strategy relies on the pre-trained NTN-DIV~\cite{xia2016modeling} model, which employs a pre-trained NTN-DIV as the evaluation model, and combines the results with MDP to produce the final ranking solution. The two strategies are, respectively, realized by the proposed MDP-DIV-kNN and MDP-DIV-NTN methods in this paper. Both methods are based on the original MDP-DIV, and they differ from each other in the sampling procedure of the episode. Suppose we are given $N$ labeled training data $D = \{ ({q^{(n)}},{X^{(n)}},{J^{(n)}})\} _{n = 1}^N$, where each query $q^{(n)}$ is associated with a set of retrieved documents $X^{(n)} = \{ {x_1^{(n)}},...,{x_M^{(n)}}\}$, and $J^{(n)}$ denotes the labels on the documents, in the form of a binary matrix. ${J^{(n)}}(i,j)=1$ if document $x_i^{(n)}$ contains the $j$-th subtopics of $q^{(n)}$ and 0 otherwise. The reward function $R({s_t},{a_t}) = \alpha \text{-} DCG[t + 1] - \alpha \text{-} DCG[t]$ is based on $\alpha$-DCG. As an overview of our approaches, we first summarize main procedure in Algorithm 1. Clearly, similar to the MDP-DIV model, our approaches also work in an iterative manner. The main improvements come from the step 4, where two different sampling methods are developed to efficiently search the action space. Next, we will elaborate the two methods.

\begin{algorithm}
\scriptsize
{\caption{MDP-DIV-kNN and MDP-DIV-NTN}}
\SetKwInOut{Input}{input}\SetKwInOut{Output}{output}
    \Input{Labeled training set $D = \{ ({q^{(n)}},{X^{(n)}},{J^{(n)}})\} _{n = 1}^N$, learning rate $\eta$, discount factor $\gamma$, reward function R, and the size of returned list m}
    \Output{All the parameters $\Theta$}
\nl    Randomly initialize $\Theta$ in $[-1,1]$\\
\nl    \While{not converge}{
\nl        \For{$(q,X,J) \in D$}{
\nl        		(${s_0},{a_0},{r_1},...,{s_{M - 1}},{a_{M - 1}},{r_M}$) $\leftarrow$ SampleEpisode($\Theta,q,X,J,R$) with kNN strategy for MDP-DIV-kNN or pre-trained NTN-DIV strategy for MDP-DIV-NTN\\
\nl        		\For{$t=0$ to $m-1$}{
\nl					${G_t} \leftarrow \sum\nolimits_{k = 0}^{M - 1 - t} {{\gamma ^k}{r_{t + k + 1}}}$\\
\nl  				$\Theta  \leftarrow \Theta  + \eta {\gamma ^t}{G_t}{\nabla _\Theta }\log \pi ({a_t}|{s_t};\Theta )$

%update parameters $\Theta$ with policy gradient algorithm of REINFORCE based on the sampled episode\\
				}
			}
        }
\nl    \Return $\Theta$
\end{algorithm}

\vspace{-3ex}

\subsection{K Nearest Neighbors Strategy}
The action evaluation is always a parameterized function that takes both state and action as input. Hence, each time to select an action, $\left| \CMcal{A} \right|$ evaluations have to be performed first, where $\left| \CMcal{A} \right|$ is the size of action space. However, this quickly becomes intractable, especially if the parameterized function is costly to evaluate. In MDP-DIV, the policy $\pi (a|s)$ is defined as a normalized softmax function whose input is the bilinear product of the utility and the selected document in Eq.(2):

\vspace{-2ex}

\begin{gather}\label{equ:equ2}
\pi ({a_t}|[{Z_t},{X_t},{h_t}]) = \frac{{\exp \{ x_{m({a_t})}^TU{h_t}\} }}{Z} \notag \\
Z = \sum\limits_{a \in A({s_t})} {\exp \{ x_{m(a)}^TU{h_t}\} }
\end{gather}

\noindent where $U$ is the parameter in the bilinear product and $Z$ is the normalization factor. The perceived utility of information $h_t$ could be computed in a recurrent manner in Eq.(\ref{equ:equ3}):

\vspace{-2ex}

\begin{equation}\label{equ:equ3}
{h_t} = \sigma (V{x_{m({a_t})}} + W{h_{t - 1}})
\end{equation}

\noindent where $V$ is the document-state transformation matrix that adds the newly perceived utility from the recently-selected document. $W$ is the state-state transformation matrix which determines the utility remained across time step. Generally speaking, at each time step, the utility perceived by users for fulfilling the information needs has to take all the previously selected documents into account, i.e., the later, the more complicated. Unfortunately, the execution complexity grows quadratically with $\left| \CMcal{A} \right|$ which makes this approach inefficient. This motivate us to reduce the computational complexity.

Since the complexity of MDP-DIV closely relates to $\left| \CMcal{A} \right|$, it is natural to find a way to ``shrink" the action space, i.e. reduce the complexity. To maintain the accuracy not degrading, the ``shrink'' strategy guarantees such foundations that: (i) It has the ability to smartly prune part of the redundant (highly similar) actions; (ii) The shrunken action evaluation can nearly generalize over actions. For search result diversification, our goal is to return the most relevant documents to the queries and ensure the diversity of the documents simultaneously. Therefore, consider such a situation: $a_i$ and $a_j$ are highly alike and both are closely relevant to the queries, can we just return $a_i$ (or $a_j$)? The answer is positive, because learning about $a_i$ also inform us about $a_j$. Moreover, in order to guarantee the diversity of the selected documents, returning them both is not a reasonable choice. Therefore, we propose a $k$ nearest neighbor based strategy (MDP-DIV-kNN) to reduce the complexity of MDP-DIV. The basic idea of the MDP-DIV-kNN is to discard the $k$ nearest neighbors of the recently-selected action (document) at each time step. In particular, the strategy is instantiated in Algorithm 2. Each time we adopt an action $a_t \in A(s_t)$, at the same time, we remove the $k$ nearest neighbors of $a_t$ from the action space, where the neighbors are computed by using the document embeddings\footnote{All the queries and documents are embedded with doc2vec~\cite{pmlr-v32-le14} embedding model.} with Euclidean distance as:

\begin{equation}\label{equ:equ4}
{f_k}({a_t}) = \mathop {\arg \min }\limits_{a \in A({s_t})}^k {\left\| {{a_t} - a} \right\|_2}
\end{equation}

The kNN lookup is a lightweight operation than the action evaluation execution although they are of the same complexity of the action space. Therefore, the kNN based strategy offers us three merits here: (i) It provides sub-quadratic complexity with respect to the action space; (ii) It avoids heavy cost of evaluating all actions while retraining generalization over actions; (iii) It directly optimizes the diversity of the selected documents.

\vspace{-3ex}

\begin{algorithm}
\scriptsize
{\caption{SampleEpisode with kNN strategy}}
\SetKwInOut{Input}{input}\SetKwInOut{Output}{output}
    \Input{$\Theta, q, X, J$, R, and m}
    \Output{An episode}
\nl    	Initialize $s_0$ and E=()\\
\nl    	\For{$t=0$ to $m-1$}{
\nl    		sample $a_t \in A(s_t)$ according to $\pi(a_t|s_t; \Theta)$\\
\nl     	$r_{t+1}=R(s_t,a_t)$\\
\nl    		change $s_t$ to $s_{t+1}$ according to the transition function\\
\nl         discard $k$ nearest neighbor of $a_t$ in $X_t$ according to Eq.(\ref{equ:equ4})\\
\nl 		append ($s_t,a_t,r_{t+1}$) to the tail of E\\
       }      
\nl    \Return E;
\end{algorithm}

\vspace{-7ex}

\subsection{Pre-trained NTN-DIV Strategy}
The other method we propose to speed up the convergence rate of MDP-DIV is to use a pre-trained diversity ranking model. As aforementioned that the large action space and the data scarcity will lead to low convergence rate. The proposed $k$ nearest neighbors strategy in turn reduces the action space at each position by filtering out the $k$ nearest neighbors of the recently-selected action (document). It is apparent that this strategy will efficiently shrink the action space to speed up the convergence. However, it cannot deal with the data scarcity. Because, in the incipient phase, once the document is selected, we will delete the $k$ nearest neighbors of the selected document, but we cannot make sure that it is relevant to the query or is the right one to rank at the current position. To deal with this problem, we propose the MDP-DIV-NTN method, which has two instantiations, i.e., MDP-DIV-NTN(D) and MDP-DIV-NTN(E), to promote the performance of MDP-DIV.

The first instantiation adopts the pre-trained NTN-DIV model to rank the candidate set first and then takes actions in part of the pre-ranked list by applying the MDP-DIV. As a result, the action space is reduced as the NTN-DIV model can provide accurate candidates with good diversity. The MDP-DIV-NTN(D) offers us two merits: (i) It directly shrinks the candidate set, i.e., the action space in MDP-DIV; (ii) It straightforwardly takes out part of the irrelevant documents (the documents with none subtopics). Although the MDP-DIV-NTN(D) methods is effective, it may loss a bit of information because the NTN-DIV model is indeed not perfectly accurate.

The second variant is more precise. We utilize the pre-trained NTN-DIV model \textbf{at each position}, i.e., each time to adopt an action. Similar to kNN strategy, we summarize its sampling strategy in Algorithm 3. It can be seen that, at each step time of the training, once the agent chooses an document, we utilize the pre-trained NTN-DIV model to find the documents which are novelty to the previously selected documents and relevant to the query simultaneously. For the next time step, the agent only needs to learn on the filtered the candidate set. Moreover, this approach also provides more considerable advantages: (i) It precisely shrinks the action space; (ii) It accurately takes out the irrelevant documents.

\vspace{-3ex}

\begin{algorithm}
\scriptsize
{\caption{SampleEpisode with pre-trained NTN-DIV strategy}}
\SetKwInOut{Input}{input}\SetKwInOut{Output}{output}
    \Input{$\Theta, q, X, J$, R, m, K, and pre-trained NTN-DIV model}
    \Output{An episode}
\nl    	Initialize $s_0$, ${\hat D=()}$\{empty set of selected docs\}, and E=()\{empty episode\}\\
\nl    	\For{$t=0$ to $m-1$}{
\nl    		sample $a_t \in A(s_t)$ according to $\pi(a_t|s_t; \Theta)$ and add $a_t$ to D\\
\nl     	$r_{t+1}=R(s_t,a_t)$\\
\nl    		change $s_t$ to $s_{t+1}$ according to the transition function\\
\nl         rank the documents in $X_t$ with D and the pre-trained NTN-DIV model\\
\nl 		choose the first K documents of $X_t$ as $X_{t+1}$\\
\nl 		append ($s_t,a_t,r_{t+1}$) to the tail of E\\
       }      
\nl    \Return E;
\end{algorithm}

\vspace{-4ex}

However, the training of the NTN-DIV model using the original implementation is time consuming\footnote{https://github.com/sweetalyssum/DiverseNTN}, because it is executed sequentially on CPU. In order to accelerate the training, we re-implement this model with Tensorflow~\cite{abadi2016tensorflow} on a NVIDIA$^{\circledR}$ Tesla$^{\circledR}$ K80 GPU because all the tensor product can be computed parallelly. Finally, we obtain a slightly better performance with less than 30 minutes to train instead of more than 5 hours training of the original CPU version. We also note that the NTN-DIV is trained off-line and its GPU implementation brings no improvement on the convergence for the MDP-DIV-NTN.

\section{Experimental Study}

\subsection{Datasets and Evaluation Metrics}

The dataset is provided by the authors\footnote{The datasets and source code are available at https://github.com/sweetalyssum/RL4SRD} which is a combination of four TREC benchmark datasets: TREC 2009-2012 Web Track. The retrieved documents are carried out on the ClueWeb09 Category B data collection\footnote{http://lemurproject.org/clueweb09/}, which is comprised of 50 million English web documents. We note that the large number of parameters in MDP-DIV needs lots of labeled data to train, which is the reason why the four benchmark datasets are merged together. In total, there are 200 queries. Each query includes several subtopics identified by the TREC assessors. Moreover, the documents' relevance labels are made at the subtopic level, which are binary with 0 denoting irrelevant and 1 denoting relevant.

We employ three widely-used evaluation metrics to assess the diverse ranking models. They are $\alpha$-NDCG~\cite{clarke2008novelty}, subtopic recall~\cite{zhai2003beyond} (denoted as ``S-recall"), and ERR-IA~\cite{chapelle2011intent}. The $\alpha$-NDCG and ERR-IA adopt the default settings in official TREC evaluation program\footnote{http://trec.nist.gov/data/web/12/ndeval.c}, which measure relevance and diversity of the ranking list by explicitly rewarding diversity and penalizing redundancy observed at each rank. The parameter $\alpha$ in these two evaluation metrics are set to 0.5. The traditional diversity metric S-recall measures the coverage rate of the retrieved subtopics for each query. All of the measures are computed over the top-$k$ search results ($k=5$ and $k=10$).

\subsection{Experimental Setup}

All the experiments are conducted with 5-fold cross-validation. We randomly re-split the queries into five even subsets\footnote{The authors does not provide the split results, therefore we re-split the queries.}. For each fold, three subsets are utilized for training, one is for validation, and the rest one for testing. Moreover, for fair comparison, we run each fold five times, and the results reported are presented with average and standard deviation values over the total 25 trials. All the experiments are performed on an intel$^{\circledR}$ Xeon$^{\circledR}$ Processor E5 V4 server with NVIDIA$^{\circledR}$ Tesla$^{\circledR}$ K80 GPU and over 256 GB memory.

We compare the proposed methods with the latest state-of-the-art baselines in search result diversification, including the NTN-DIV~\cite{xia2016modeling} and MDP-DIV~\cite{Xia:2017:AMD:3077136.3080775}. We do not compare conventional models because previous studies have shown that their performances are inferior~\cite{xia2016modeling,Xia:2017:AMD:3077136.3080775}.

\textbf{NTN-DIV}: As mentioned in Section 3.3, as a state-of-the-art method, the model computes a ranking score by taking both relevance and novelty into account with a neural tensor network. To speed up the training, we implement this method with Tensorflow on GPU which is extremely much faster than the original CPU version. The tensor slices is 100.

\textbf{MDP-DIV}: As introduced in Section 3.2, this is the latest and state-of-the-art method based on the MDP. We set parameters following~\cite{Xia:2017:AMD:3077136.3080775}, because the datasets utilized are exactly the same as in~\cite{Xia:2017:AMD:3077136.3080775}. As our methods employ $\alpha$-DCG as reward function, the $\alpha$-DCG version MDP-DIV is thus adopted for a fair comparison.

\textbf{MDP-DIV-kNN}: The parameter $k$ is set to be $10\% \times \left| \CMcal{A} \right|$, $20\% \times \left| \CMcal{A} \right|$, and $30\% \times \left| \CMcal{A} \right|$, denoted as MDP-DIV-kNN(10), MDP-DIV-kNN(20), and MDP-DIV-kNN(30), respectively. The other parameters follow the settings in MDP-DIV.

\textbf{MDP-DIV-NTN}: The tensor slices of the pre-trained NTN-DIV model is 100, and the learning rate is 0.009. The size of both pre-ranked list in MDP-DIV-NTN(D) and MDP-DIV-NTN(E) is set to $50\% \times \left| \CMcal{A} \right|$. Again, the other parameters follow the setting in MDP-DIV.

In the experiments, the query vector and document vector are represented as the embeddings generated by the Doc2vec model, which is trained on all the documents in Web Track datasets. When training of the Doc2vec model, the number of dimension is set to 100, the learning rate is set to 0.025 and 8 is utilized as the window size.

\subsection{Results and Analysis}

\vspace{-3ex}

\begin{table}[ht]
  \centering
  \scriptsize
  %\vspace{-3ex}
  \caption{Performance comparison of all methods on TREC web track dataset. (The best results are marked in bold format)}
  \begin{tabular}{l|ccccccc}
    \toprule
    Method                   & $\alpha$-NDCG@10 & $\alpha$-NDCG@5  & S-recall@10 & S-recall@5  & ERR-IA@10  & ERR-IA@5 & time (:h)\\
    \midrule
    NTN-DIV(GPU)             & 0.4617           & 0.4124           & 0.6205      & 0.5140      & 0.3446     & 0.3186   & \textbf{0.5}\\
    MDP-DIV                  & 0.4874           & 0.4480           & 0.6639      & 0.5599      & 0.3697     & 0.3477   & 65 \\
    \midrule
    MDP-DIV-kNN(10)          & 0.4915           & 0.4462           & 0.6731      & 0.5435      & 0.3725     & \textbf{0.3539}   & 43 \\
    MDP-DIV-kNN(20)          & 0.4869           & 0.4461           & 0.6582      & 0.5463      & 0.3723     & 0.3506   & 25 \\
    MDP-DIV-kNN(30)          & 0.4844           & 0.4464           & 0.6489      & 0.5467      & 0.3721     & 0.3517   & 16 \\
    \midrule
    MDP-DIV-NTN(D)           & 0.4912           & 0.4470           & 0.6738      & 0.5464      & 0.3727     & 0.3493   & 26\\
    MDP-DIV-NTN(E)           & \textbf{0.4937}  & \textbf{0.4485}  & \textbf{0.6795} & \textbf{0.5627} & \textbf{0.3735} & 0.3497   & 53 \\
    \bottomrule
  \end{tabular}
  \label{tab:table1}
\end{table}

\vspace{-3ex}

\textbf{Performance Comparison for Search Result Diversification.} Table \ref{tab:table1} shows the performance of all the methods on TREC web track datasets. From the table, we can see that the re-implemented GPU version of NTN-DIV needs half an hour to train which is extremely faster than all the other methods. However, its performance (accuracy) is significantly inferior to the other approaches. 

Compared to the original MDP-DIV, the proposed MDP-DIV-kNN methods and MDP-DIV-NTN methods are all faster, with a barely degraded or even slightly better accuracy. Among the MDP-DIV-kNN methods, the fastest one is the MDP-DIV-kNN(30) which discards $30\%$ of the current actions by the nearest neighbor strategy. It takes 16 hours to train which is 3x faster than the MDP-DIV (taking 65 hours). Moreover, the MDP-DIV-kNN(10) shows best accuracy among the three. We observe that it is slightly better than the original MDP-DIV, while the other two (i.e., MDP-DIV-kNN(20) and MDP-DIV-kNN(30)) are slightly worse. The reasons are two fold: (i) The $k$ nearest neighbors strategy can help produce a more diverse ranking list; (ii) Filtering nearest neighbors may also result in a information loss. The lager the $k$, the more the information loss is. Therefore, the performance is a trade-off between the complexity and the accuracy. 

As to the MDP-DIV-NTN methods, the performance is better compared to MDP-DIV. For MDP-DIV-NTN(D), the pre-tained NTN-DIV model offers us a pre-ranked list which helps to shrink the action space and filters part of the irrelevant document; For MDP-DIV-NTN(E), at each time step, we model the novelty of the candidate document based on both the query and preceding selected documents which provides us a more accurate pre-ranked list. Hence, its performance (accuracy) is not only better than MDP-DIV, but also better than MDP-DIV-NTN(D). However, the computing on GPU at each time step will cost some time. This is the reason that MDP-DIV-NTN(E) (taking 53 hours) does not run as fast as MDP-DIV-NTN(D) (taking 26 hours).

%\vspace{-3ex}

\begin{figure}[t]
  \centering
  \includegraphics[height=4.5cm, width = 1\textwidth]{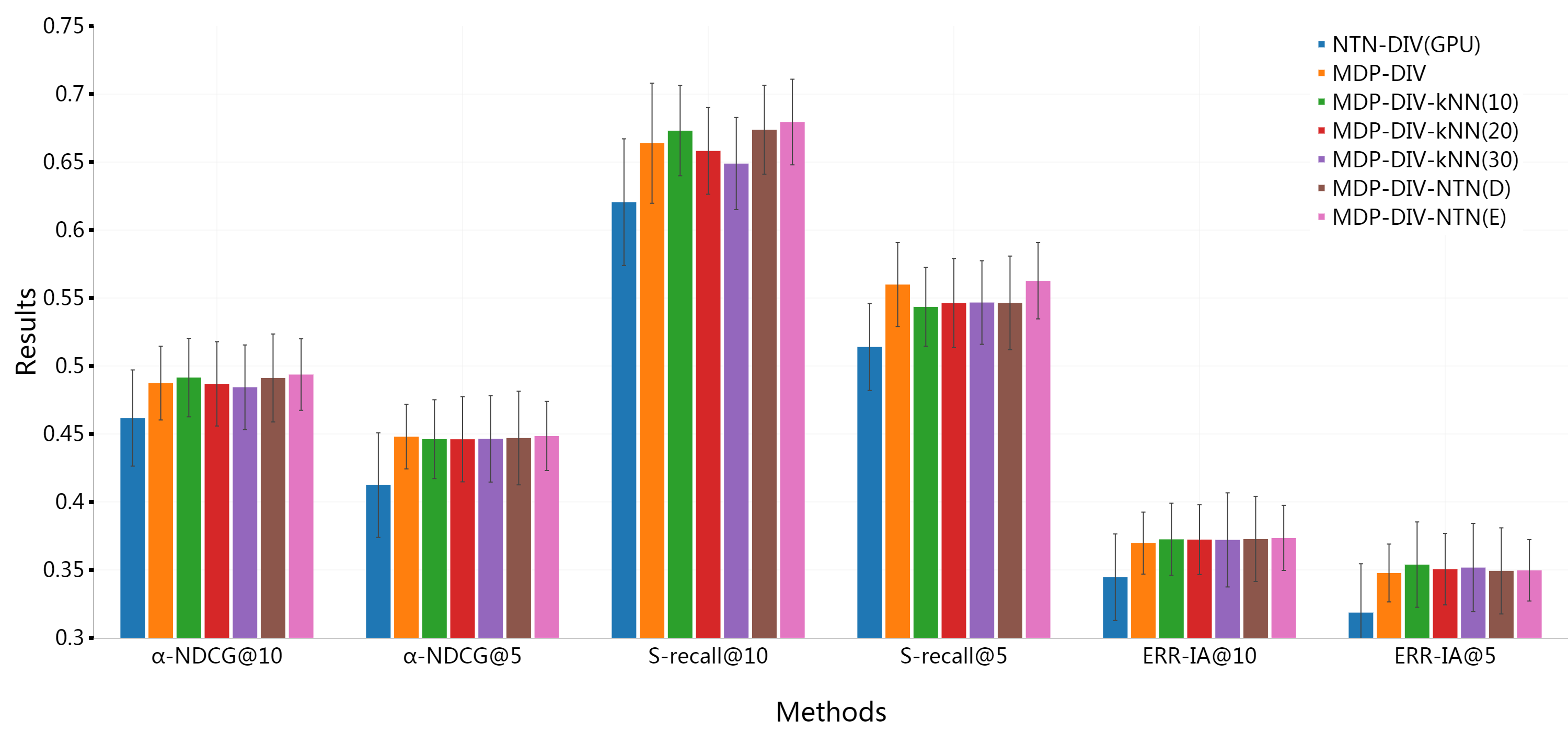}
\caption{Performance of stability comparison of all the methods on TREC web track dataset.}
\label{fig:error-bar}
\end{figure}

%\vspace{-2ex}

In Figure \ref{fig:error-bar}, we report the error-bar of the comparison methods. From the figure, we can see that all the approaches show relatively consistent standard deviation, which indicates the proposed methods achieve stably better or comparable performance than NTN-DIV and MDP-DIV.

Next, we present some results to analyze the efficiency and effectiveness of the proposed methods in details.

%\vspace{-3ex}

\begin{figure}[t]
  \centering
  \includegraphics[height=7.5cm, width=0.65\textwidth]{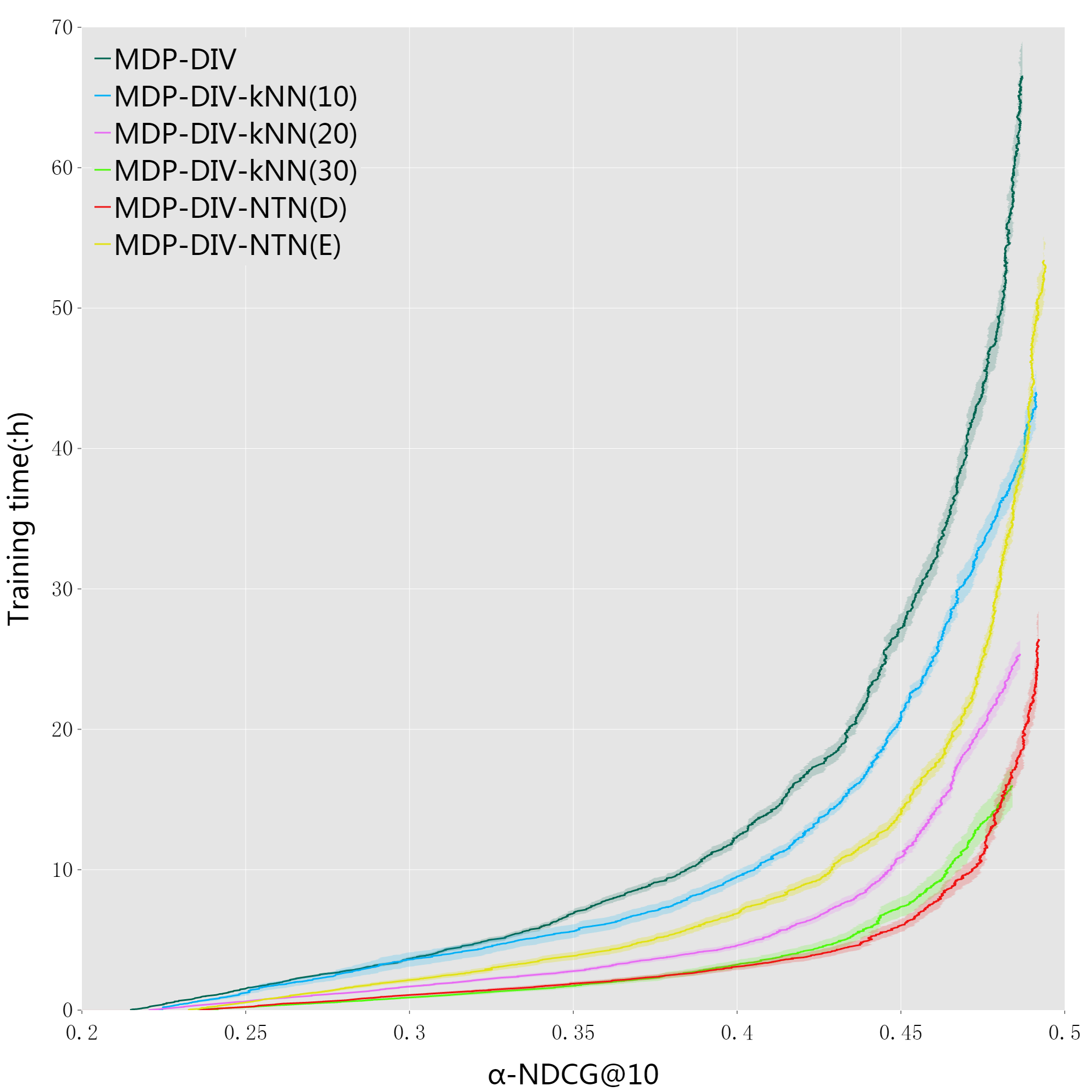}
\caption{Efficiency analysis of the proposed methods on TREC web track dataset.}
\label{fig:shadedline-1}
\end{figure}

%\vspace{-2ex}

\textbf{Efficiency Analysis.} To analyze the efficiency, We draw a shaded-line figure in Figure \ref{fig:shadedline-1} to show the time cost for $\alpha$-NDCG@10 performance of the models based on 5-fold cross validation. The horizon axis is the $\alpha$-NDCG@10 performance, and the vertical axis is the time cost to achieve the $\alpha$-NDCG@10 performance. The curve in the figure means the average time cost for $\alpha$-NDCG@10 performance, and the shade is the standard deviation. From the figure we can see that the proposed MDP-DIV-kNN and MDP-DIV-NTN methods are all trained faster than the original MDP-DIV. Specially, with the increase of the $k$ value, the MDP-DIV-kNN models converge faster. Although the accuracy of the final convergence will sacrifice, it is still relatively acceptable. The MDP-DIV-NTN(D) trained faster than other models before the $\alpha$-NDCG@10 performance reaches 0.48. However, the promotion of $\alpha$-NDCG@10 after 0.48 becomes very time-consuming. In terms of $\alpha$-NDCG@10, after convergence, MDP-DIV-NTN(D) performs worse than MDP-DIV-NTN(E) which achieves the best accuracy. 

Compared to the original MDP-DIV, for instance, to achieve the $\alpha$-NDCG@10 performance at 0.48, MDP-DIV-kNN(30) and MDP-DIV-NTN(D) are almost 3 times faster, MDP-DIV-kNN(20) is 1.4 times faster, MDP-DIV-kNN(10) is 0.4 times faster, and MDP-DIV-NTN(E) is 0.54 times faster than MDP-DIV. According to the observations, we draw the following conclusions: (i) The proposed MDP-DIV-kNN and MDP-DIV-NTN models fulfill the target of accelerate the convergence of MDP-DIV without much accuracy sacrifice; (ii) The MDP-DIV-kNN methods converge fast with a relatively acceptable accuracy, and the MDP-DIV-NTN methods converge fast and show better accuracy than MDP-DIV.

%\vspace{-3ex}

\begin{figure}[t]
  \centering
  \includegraphics[height=4.5cm, width=0.95\textwidth]{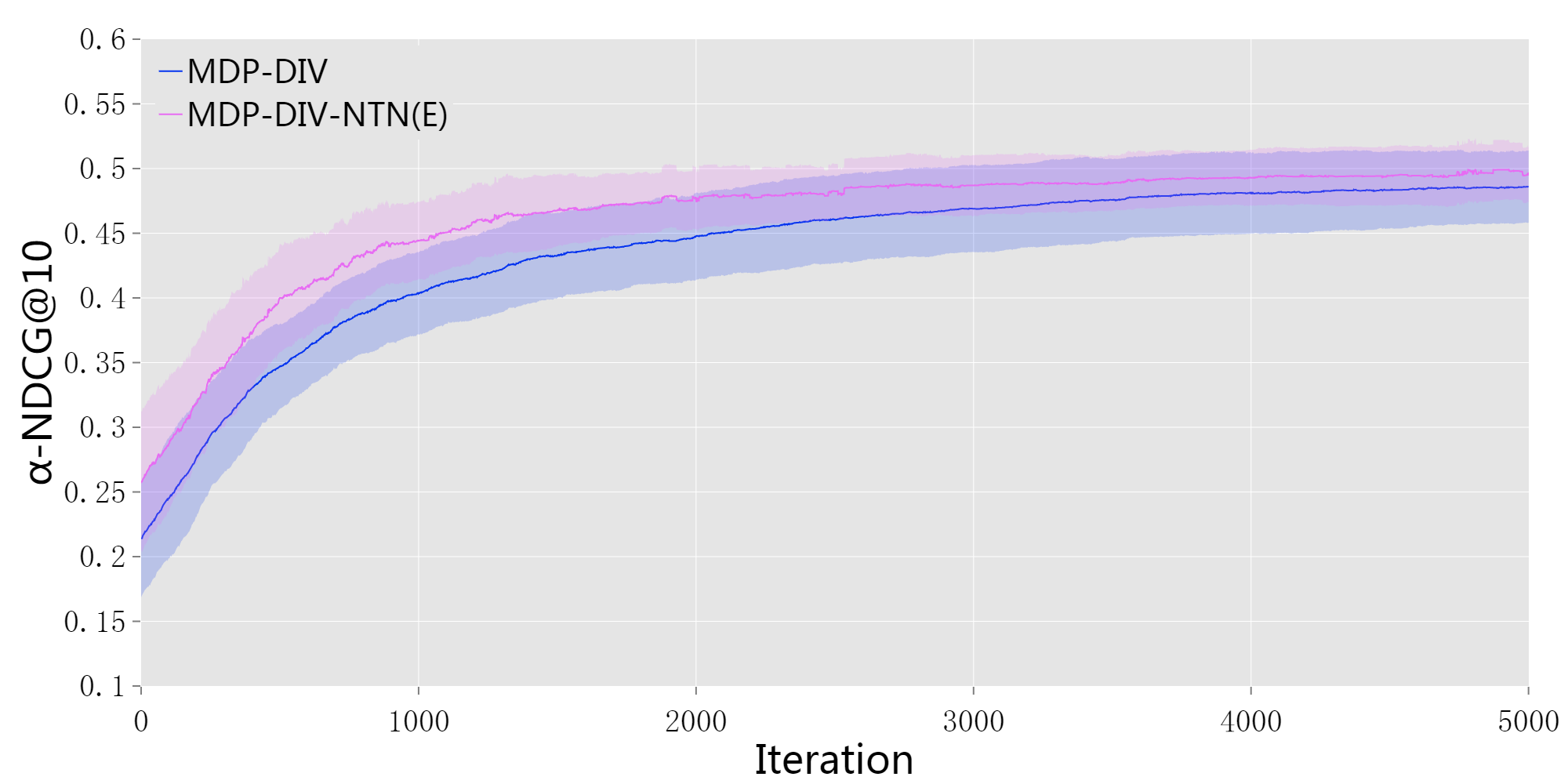}
  \caption{Effectiveness analysis of the proposed methods on TREC web track dataset.}
\label{fig:shadedline-2}
\end{figure}

%\vspace{-3ex}

\textbf{Effectiveness analysis.} Another promotion comes from the accuracy performance. Here we draw a shaded-line figure in Figure \ref{fig:shadedline-2} to show the $\alpha$-NDCG@10 performance against the number of iterations. From this figure, we observe that during the first 2000 iterations, MDP-DIV-NTN(E) shows a significant improvement of $\alpha$-NDCG@10 up to 0.05 over the MDP-DIV. As the training phase goes on, the improvement becomes gentle. Finally, when both the methods converge, MDP-DIV-NTN(E) still delivers better performance than MDP-DIV. In summary, we draw the following conclusions: (i) The proposed MDP-DIV-NTN(E) converges faster than the original MDP-DIV; (ii) MDP-DIV-NTN(E) can reach a high performance in the first 2000 iterations, and the converge performance is also better. The reason of the fast convergence rate is that we utilize an off-line NTN-DIV model to shrink the search space and filter part of the irrelevant documents.

\section{Conclusion}
In this paper, we aim to promote the performance of MDP-DIV by speeding up its convergence rate without much accuracy sacrifice. After analysis, we find the slow convergence of MDP-DIV is mainly due to the two reasons: the large action space and data scarcity. On the one hand, the sequential decision making at each position needs evaluate the query-document relevance for all the candidate set, which results in a huge searching space for MDP; on the other hand, due to the data scarcity, the agent has to proceed more ``trial and error" interactions with the environment. To tackle this problem, we propose MDP-DIV-kNN and MDP-DIV-NTN methods. The experiment results demonstrate that the two proposed methods indeed accelerate the convergence rate of the MDP-DIV, while the accuracies produced barely degrade, or even become better.

\textbf{Acknowledgement.} This research was supported in part by NSFC under Grant Nos. 61602132 and 61572158, and Shenzhen Science and Technology Program under Grant No.JCYJ20160330163900579.

%\begin{thebibliography}{5}
%

%\end{thebibliography}

\bibliographystyle{splncs03}
\bibliography{sigproc}

%
% second contribution with nearly identical text,
% slightly changed contribution head (all entries
% appear as defaults), and modified bibliography
%

\end{document}